\documentclass[]{emulateapj}
\usepackage{color}
\usepackage{mathtools}

\newcommand{\str}{Str\"{o}mgren }

\newcommand{\hi}{\rm H\,{\textsc i}}
\newcommand{\hii}{\rm H\,{\textsc{ii}}}
\newcommand{\myscale}{1.15}
\newcommand{\msun}{M_\sun}

\newcommand{\mbh}{M_{\rm BH}}

\newcommand{\mach}{\mathcal{M}}

\newcommand{\khp}[1]{{\color{black}#1}}

\shorttitle{Gaseous Dynamical Friction in Presence of BH Radiative Feedback}
\shortauthors{Park and Bogdanovi\'c}

\begin{document}

\title{Gaseous Dynamical Friction in Presence of Black Hole Radiative Feedback}
\author{KwangHo Park}
\author{Tamara Bogdanovi\'c}
\affil{Center for Relativistic Astrophysics, School of Physics,
Georgia Institute of Technology, Atlanta, GA 30332, USA}
\email{kwangho.park@physics.gatech.edu, tamarab@gatech.edu}

\begin{abstract}
Dynamical friction is thought to be a principal mechanism responsible
for orbital evolution of massive black holes (MBHs) in the aftermath
of galactic mergers and an important channel for formation of
gravitationally bound MBH binaries. We use 2D radiative
hydrodynamic simulations to investigate the efficiency of dynamical
friction in the presence of radiative feedback from an MBH moving through
a uniform density gas. We find that ionizing radiation that emerges from the
innermost parts of the MBH's accretion flow strongly affects the
dynamical friction wake and renders dynamical friction inefficient
for a range of physical scenarios. MBHs in this regime tend to
experience positive net acceleration, meaning that they speed up,
contrary to the expectations for gaseous dynamical friction in
absence of radiative feedback. The magnitude of this acceleration is
however negligibly small and should not significantly alter the
velocity of MBHs over relevant physical timescales. Our results
suggest that suppression of dynamical friction is more severe at
the lower mass end of the MBH spectrum which, compounded with
inefficiency of the gas drag for lower mass objects in general,
implies that $<10^7\,\msun$ MBHs have fewer means to reach the
centers of merged galaxies. These findings provide formulation for
a sub-resolution model of dynamical friction in presence of MBH
radiative feedback that can be easily implemented in large scale simulations.
\end{abstract}

\keywords{accretion, accretion disks -- black hole physics --
hydrodynamics -- radiative transfer}

\section{introduction}\label{sec:intro}

A massive perturber moving through some background medium creates in
it a density wake. The wake trails the perturber and exerts on it a 
gravitational force in the direction opposite to the 
motion, thus acting as a brake and earning this interaction a name:
``dynamical friction." In his seminal work, \citet{chandra43} studied
the efficiency of dynamical friction for a massive perturber moving
through a stellar background and found the drag force to be maximized
when the velocity of the perturber is comparable to the velocity
dispersion of stars, $v\approx \sigma$. \citet{ostriker99} later
evaluated the dynamical friction force acting on a perturber traveling
through a uniform gaseous medium and found that the gaseous drag
is more efficient than stellar drag for transonic perturbers. This
work also established that the gaseous drag is operating most
efficiently when $1 < \mathcal{M} \lesssim {\rm few}$, where the
Mach number $\mathcal{M} \equiv v/c_{s,\infty}$ is defined as the
ratio between the perturber velocity $v$ and the sound speed of the
ambient gas ``at infinity" $c_{s,\infty}$ , i.e., in a distant
region unaffected by the perturber. In this mildly supersonic
regime, the dynamical friction force takes form

\begin{equation}
F_{\rm DF} = - \frac{4\pi(GM_{\rm BH})^2 \rho_\infty}{v^2}
\ln{\left[\Lambda \left( 1-\frac{1}{\mathcal{M}^2}\right)^{0.5}\right]} 
\label{eq_Fdf}
\end{equation}

where $M_{\rm BH}$ is the mass of the perturber (hereafter assumed
to be a black hole; BH) and $\rho_\infty$ is the density of the
gas at infinity. $\ln \Lambda= \ln{(r_{\rm max} / r_{\rm min})}$
is the Coulomb logarithm and $r_{\rm min}$ and $r_{\rm max}$ represent
the smallest and largest spatial scales in the gas wake that
contribute to dynamical friction, respectively. Based on
Equation~(\ref{eq_Fdf}), it follows that the effect of dynamical
friction is expected to be stronger for massive perturbers given
that they can interact with a sufficiently dense pool of ambient
gas.

Because of its efficiency, gaseous drag has been extensively
investigated in simulations that follow pairing of massive black
holes (MBHs) in gas rich mergers of galaxies \citep[see][for
review]{mayer13}. The most important questions for this research
area are: in which galaxies does gaseous dynamical friction lead to
successful gravitational pairing of MBHs and on what timescales?
\cite{callegari09,callegari11}, for example, find that MBH pairs with
mass ratios $q<0.1$ are less likely to form gravitationally bound
binaries within a Hubble time, largely due to the inefficiency of
the gas drag on the smaller of the two MBHs. When pairing is
successful, gaseous dynamical friction is capable of transporting
the MBHs from galactic radii of a few hundred to $\sim1$\,pc on
timescales as short as $\sim10^7$\,year and tens of times faster than
stellar dynamical friction \citep[for e.g.,][]{escala05, dotti06,
mayer07}.

Interestingly, gravitational interaction of the MBH with the
surrounding gas is quite local. For example, for a BH with mass $\sim
10^6\,\msun$ most of the gravitational drag force is contributed
by the gas that resides within only a few parsecs of the MBH
\citep{chapon13}. This proximity implies that the dynamical friction 
wake can be strongly affected, and possibly obliterated, by irradiation
and feedback from an accreting MBH. \khp{Indeed, studies of dynamical
evolution of MBHs find that dynamical friction can be significantly
reduced due to the wake dispersal caused by purely thermal feedback
from MBHs in simulations that follow gravitationally recoiled MBHs
\citep{Sijacki:11} and BH pairing \citep{SouzaLima:2016}
where the term ``wake evacuation" is dubbed.} These results
bring into question the assumed efficacy of gaseous dynamical
friction and call for further exploration of the effects of radiative
feedback powered by MBH accretion.

Here, we investigate this question by considering interaction of 
matter and radiation in the gravitational
potential well of a moving MBH. The most important findings of this
work are that there are regimes, set by the properties of the MBH
and its ambient medium, in which gas dynamical friction is rendered
inefficient by the MBH feedback. We lay out the relevant physical
regimes and scales in Section~\ref{sec:regimes} and evaluate the
efficiency of dynamical friction using local radiation hydrodynamic
simulations in Section~\ref{sec:method}. We discuss implications
of our findings and conclude in Section~\ref{sec:conclusions}.

\begin{table*}
\begin{center}
\caption{Simulation Parameters}
\begin{tabular}{cccccccccc}
\hline
\hline
ID & $N_{\rm r}\times{N}_{\theta}$ & $\Delta r/r$ & $\mach$ & $t_{\rm end}\,$(Myr)  \\
\hline
HR & 400$\times$128 & 0.026 & 0.5, 1.0, 2.0 & 201.0, 74.9, 74.9
\\
MR & 300$\times$96 & 0.035  & 0.5, 1.0, 2.0
& 201.0, 201.0, 201.0  \\
LR & 200$\times$64 & 0.053  & 0.5, 1.0, 1.5,
2.0 & 201.0, 201.0, 201.0, 201.0 \\
\hline
\end{tabular}
\label{table:para}
\end{center}
\end{table*}

\section{Regimes of dynamical friction in the presence of radiative feedback}\label{sec:regimes}


Photoionization and radiation pressure exerted by radiation escaping
from the innermost parts of the BH accretion flow can create the
ionized region (the Str\"omgren sphere or \hii~region) around the
a BH and strongly alter the properties of surrounding gas. This
radiation ``response", also referred to as the BH radiative feedback,
has been found to lower the accretion
rate by orders of magnitude relative to a fiducial case in which
radiative feedback is neglected \citep{MiloCB:09,Li:11,ParkR:11,ParkR:12,
PacucciF:2015, Park:2016}. In the context of dynamical friction,
this accretion regime is also likely to correspond to reduced
efficiency of the gas drag due to the impact of ionizing radiation
on the BH's density wake.

\citet{Park:2014a} and \cite{Inayoshi:2016} however find a physical 
regime at higher gas densities in which the nature of accretion onto 
the BH fundamentally changes from this picture, as the \hii~region
collapses under the gravity of the surrounding gas, and accretion continues
unaffected by radiation at supper Eddington rates. The two studies
differ in their treatment of radiation transport: the latter considers
the effects of radiation trapping at high accretion rates
\citep{Begelman:78, Begelman:79} while the
former neglects them. Both nevertheless indicate that conditions
for hyper-accretion arise for a BH immersed in neutral gas with
number density $n_\infty$, when $n_\infty\,M_{\rm BH} > 10^9 -
10^{10} \msun\,{\rm cm^{-3}}$. This finding implies that the density
distribution of gas in the hyper-accretion regime is only weakly
affected by radiation pressure, and thus the efficiency of dynamical
friction is likely to be restored to that predicted classically,
i.e., in absence of radiative feedback.

For moving BHs, accretion rate, and thus accretion luminosity,
also depend on BH velocity. We draw on the results of \citet[][hereafter
PR13]{ParkR:13}, who used radiation hydrodynamic simulations to
investigate the growth and luminosity of BHs moving through a uniform
gaseous medium with temperature $T_\infty = 10^4$\,K. PR13 show
that in this case the BH radiative feedback causes a formation of
an \hii~region elongated along the direction of the BH motion and
filled with ionized hydrogen (or H--He) gas of temperature, $T_{\rm
in} \approx 4\times 10^4$\,K ($6\times 10^4$\,K). A shell of gas
with increased density, corresponding to $(1+\mathcal{M}^2)\,
n_\infty$, forms in front of the BH as a consequence of the ``snowplow"
effect caused by radiation pressure. For a wide range of simulated
gas densities the shell becomes gravitationally unstable and collapses
when $\mathcal{M} \gtrsim 4$, restoring the properties of the gas
flow around the BH to the classical Bondi--Hoyle--Lyttleton solution
\citep{HoyleL:39,BondiH:44}. While PR13 have not explicitly
investigated the properties of the gas wake in their simulations,
it follows that in this regime the efficiency of dynamical friction
is restored to its classically derived value.

We therefore focus on scenarios described by $1 \le \mathcal{M} <
4$, when gas dynamical friction is classically expected to be most
efficient, and $(1+\mathcal{M}^2)\, \mbh\,n_\infty < 10^9\,
\msun\,{\rm cm^{-3}}$, when the influence of the BH radiative
feedback on dynamical friction wake is expected to be significant.
Having established the relevant regime we estimate the extent
of the dynamical friction wake and compare it to the size of the
\hii~region.

\subsection{Relevant scales}

The extent of the dynamical friction wake that would form behind the MBH in absence of radiative feedback can be estimated as a radius of the gravitational influence of the MBH
\begin{equation}
R_{\rm DF} \approx \frac{GM_{\rm BH}}{c_{\rm s,\infty}^2} =
52.1\,{\rm pc} \left(\frac{T_{\infty}}{10^4\,{\rm K}}\right)^{-1}  \left(\frac{M_{\rm BH}}{10^6\,M_\odot} \right).
\label{eq:Rw}
\end{equation}
Here $c_{s,\infty} = \sqrt{{\gamma k_{\rm B} T_\infty / \mu m_p}} = 9.1\,{\rm km\,s^{-1}} (T_\infty/10^4\,{\rm K})^{1/2}$, assuming isothermal gas with hydrogen composition characterized by the adiabatic index $\gamma =1$ and mean atomic weight $\mu =1$. The constants have their usual meaning. As mentioned in Section~\ref{sec:intro}, most of the gravitational drag force is contributed by the gas that resides within an order of magnitude smaller region than that estimated in equation~(\ref{eq:Rw}), so this value can be considered a conservative upper limit.

The size of the elongated \hii~region sensitively depends on the MBH
accretion rate and velocity. For the purpose of the estimates
presented here, we neglect the elongation of the \hii~region and
estimate the radius of a spherically symmetric ionization sphere
that would form around a stationary MBH accreting hydrogen gas,
$R_{\rm HII} = (3\dot{N}/4\pi \alpha_{\rm rec})^{1/3} n_{\infty}^{-2/3}$
\citep{osterbrock06}. The number of ionizing photons per unit time emitted
from the innermost parts of a BH's accretion flow is
\begin{equation}
\dot{N} = \int_{\nu_0}^\infty \frac{L_\nu}{h\nu} d\nu \approx \frac{\alpha - 1}{\alpha} \left(\frac{L_{\rm bol}}{h\nu_0}\right)
\label{eq:Ndot}
\end{equation}
which is characterized by bolometric luminosity $L_{\rm bol} =
\int_0^\infty L_\nu d\nu$ and $L_\nu \propto \nu^{-\alpha}$. Here,
$\alpha =1.5$ is the spectroscopic index representative of the
spectral energy distribution (SED) of active galactic nuclei (AGN)
in the high accretion state, and $h\nu_0 = 13.6\,$eV for hydrogen.
\khp{$\alpha_{\rm rec} = 4\times 10^{-13}{\rm cm^3\,s^{-1}}$
corresponds to the Case A recombination coefficient for hydrogen
gas at the temperature $T_\infty = 10^4$\,K, used here so to match
simulations described in the next section. Note that the Case B
coefficient is a more appropriate choice for the high gas densities 
in this problem, which would result in a small change \str 
radius being proportional to $\alpha_{\rm rec}^{-1/3}$.} We consider
accretion powered BH luminosity, $L_{\rm bol} = \eta \dot{M} c^2$,
and assume radiative efficiency $\eta =0.1$
\citep{ShakuraS:73}.

\begin{figure}[t]
\epsscale{1.15} \plotone{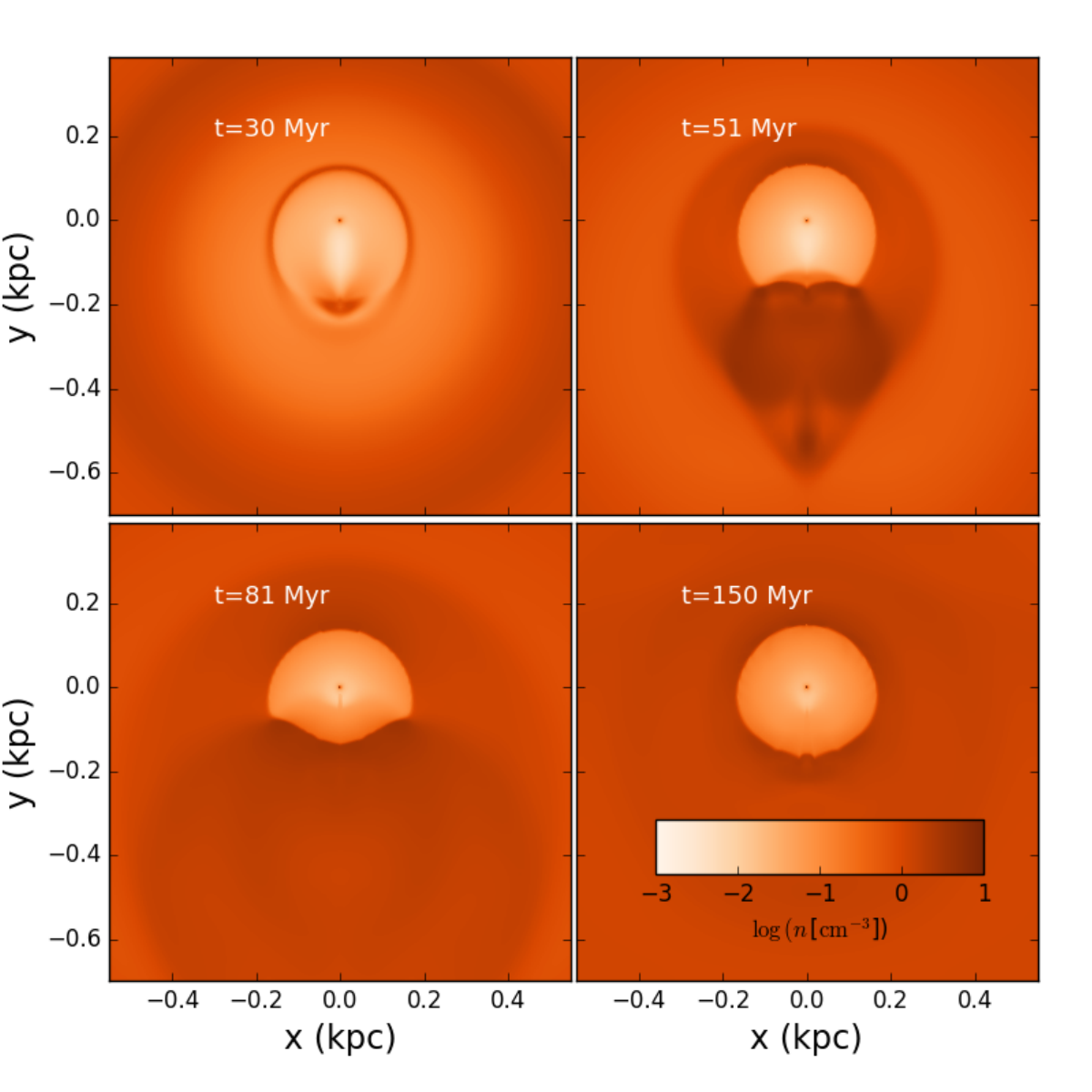}
\caption{Snapshots of gas number density at different times for the HR
run with $\mach=0.5$. The flow of gas is from top to bottom and the
BH is located at $(x,y)= (0,0)$ in each panel. The shape of the \hii~region and dynamical friction wake reach steady state after 150\,Myr
in this simulation. }
\label{fig:snapshot_m05} 
\end{figure}

These ingredients provide an estimate for the size of the \hii~region, given a known MBH accretion rate. In the case of moving MBHs with the Mach number in the range $1 < \mathcal{M} < 4$, PR13 find that the accretion rate mediated by radiation feedback can be expressed as 
\begin{equation}
\dot{M} \approx 1.2\times 10^{-2} \mathcal{M}^2  \dot{M}_{\rm B} \left(\frac{T_\infty}{T_{\rm in}} \right)^{5/2}
\label{eq:Mdot}
\end{equation}
where $\dot{M}_{\rm B}$  is the nominal Bondi accretion rate for isothermal gas \citep{BondiH:44}
\begin{equation} \begin{split}
&\dot{M}_{\rm B}= \frac{\pi e^{3/2} \rho_\infty (GM_{\rm BH})^2}{c_{s, \infty}^3} \\ 
&=8.7\!\times\!10^{-3}\!
\left(\frac{n_\infty}{1\,{\rm cm^{-3}}} \right)
\left(\frac{T_\infty}{10^4\,{\rm K}} \right)^{-\frac{3}{2}}\!
\left(\frac{M_{\rm BH}}{10^6 M_\odot} \right)^2\, 
\msun\,{\rm yr}^{-1}. 
\label{eq:M_B}
\end{split} \end{equation}

Combining Equations~(\ref{eq:Ndot})--(\ref{eq:M_B}), we obtain the estimate for the radius of the \hii~sphere
\begin{equation}
\small{
R_{\rm HII} \approx 430\,{\rm pc}\; \mathcal{M}^{2/3}
\left(\frac{n_\infty}{1\,{\rm cm^{-3}}} \right)^{-\frac{1}{3}}\!
\left(\frac{T_{\rm in}}{4\!\times\!10^4\,{\rm K}} \right)^{-\frac{1}{2}}\!
\left(\frac{M_{\rm BH}}{10^6 M_\odot} \right)^{\frac{2}{3}}.
}
\label{eq:R_HII}
\end{equation}
Considering the ratio of Equations~(\ref{eq:R_HII}) and (\ref{eq:Rw}) we get
\begin{equation}
\small{
\frac{R_{\rm HII}}{R_{\rm DF}} \sim 8 \mathcal{M}^{\frac{2}{3}}
\left(\frac{T_\infty}{10^4\,{\rm K}} \right)
\left(\frac{T_{\rm in}}{4\!\times\!10^4\,{\rm K}} \right)^{-\frac{1}{2}}
\left(\frac{\mbh n_\infty}{10^6 M_\odot\, {\rm cm^{-3}}} \right)^{-\frac{1}{3}}
}
\end{equation}
where we find that $R_{\rm HII} \gtrsim R_{\rm DF}$ for all values of the
Mach number in the range $1 < \mathcal{M} < 4$ and when
$(1+\mathcal{M}^2)\, \mbh\,n_\infty < 10^9\,M_\odot\,{\rm
cm^{-3}}$. It follows that in this regime the dynamical friction
wake is likely to be fully ionized and dispersed by the MBH radiative
feedback, especially given the proximity of the wake mentioned before.

In the next section we examine the detailed structure of the wake and
the \hii~region, and numerically evaluate the resulting dynamical
friction force from radiation hydrodynamic simulations.

\begin{figure}[t]
\epsscale{\myscale} \plotone{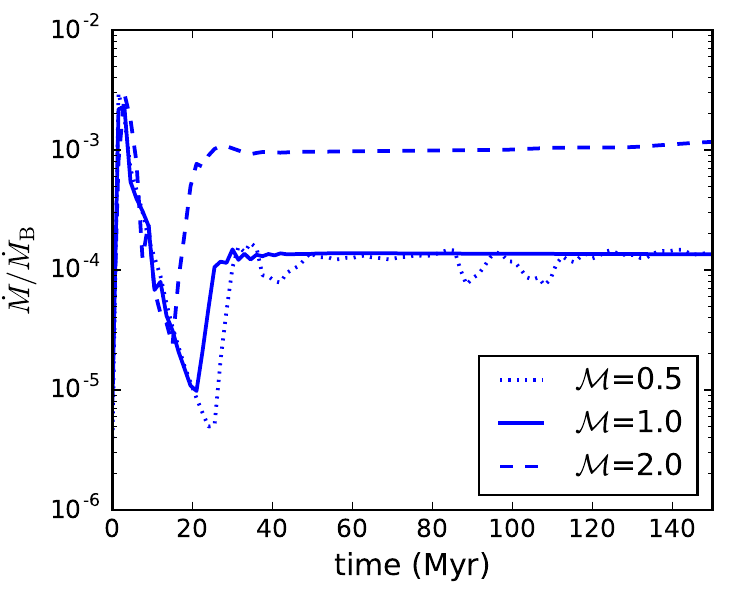} 
\caption{Accretion rate onto the MBH as a function of time shown in units of Bondi accretion rate defined in Equation~(\ref{eq:M_B}). Different line styles mark MR runs with $\mach$ = 0.5, 1.0, and 2.0. 
} 
\label{fig:accrate} \end{figure}

\begin{figure*}[t]
\epsscale{1.2} \plotone{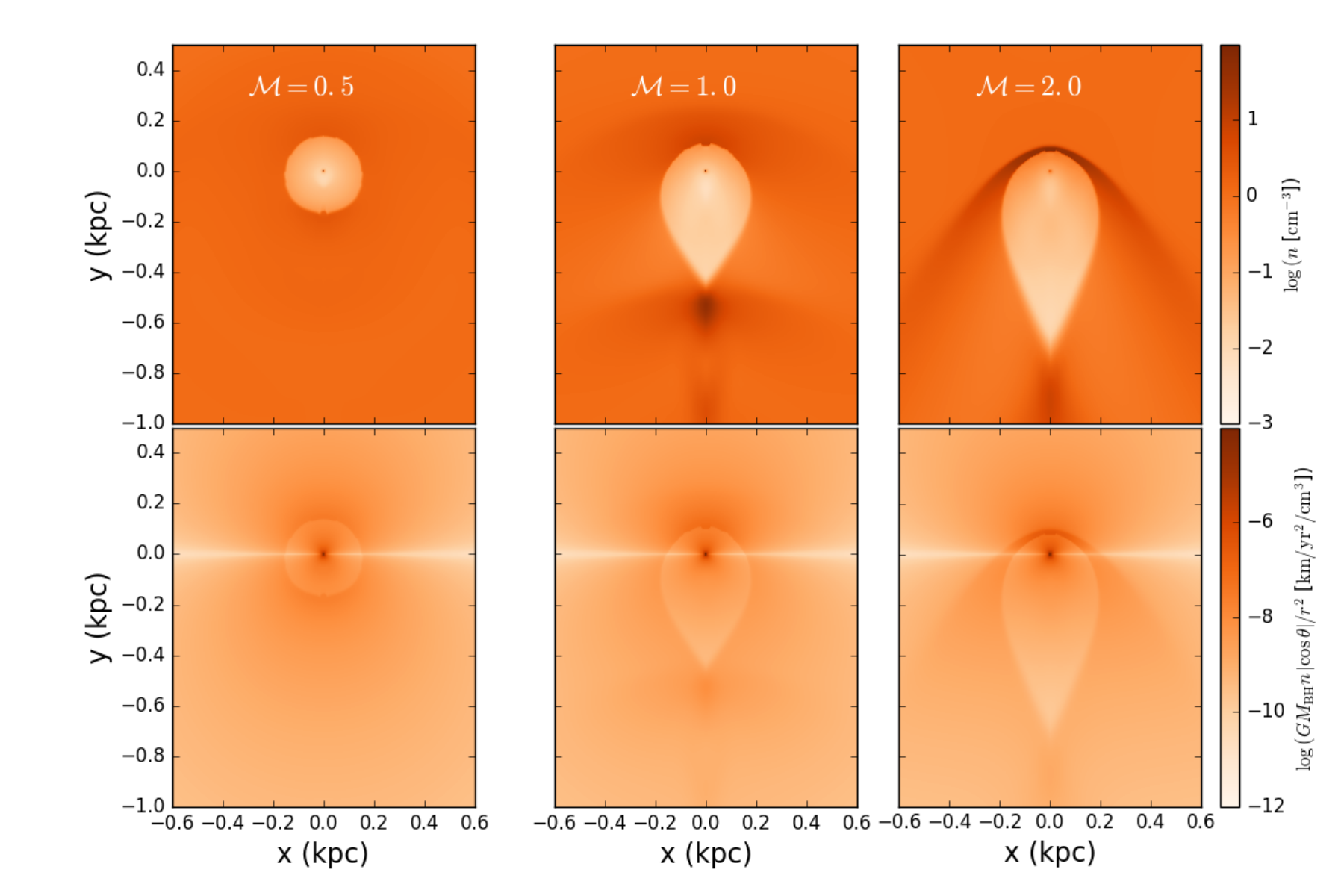} 
\caption{{Top}: 2D snapshots of the number density of gas at
t = 150\,Myr for MR runs $\mach$ = 0.5, 1.0, and 2.0 (left to right).
All snapshots
show steady state gas distributions. {Bottom}: contribution to
dynamical friction from the gas surrounding the MBH, evaluated as
the gravitational acceleration per unit volume along the direction of the
MBH motion. }
\label{fig:snapshot} \end{figure*}

\begin{figure*}[t] \epsscale{1.15}
\plottwo{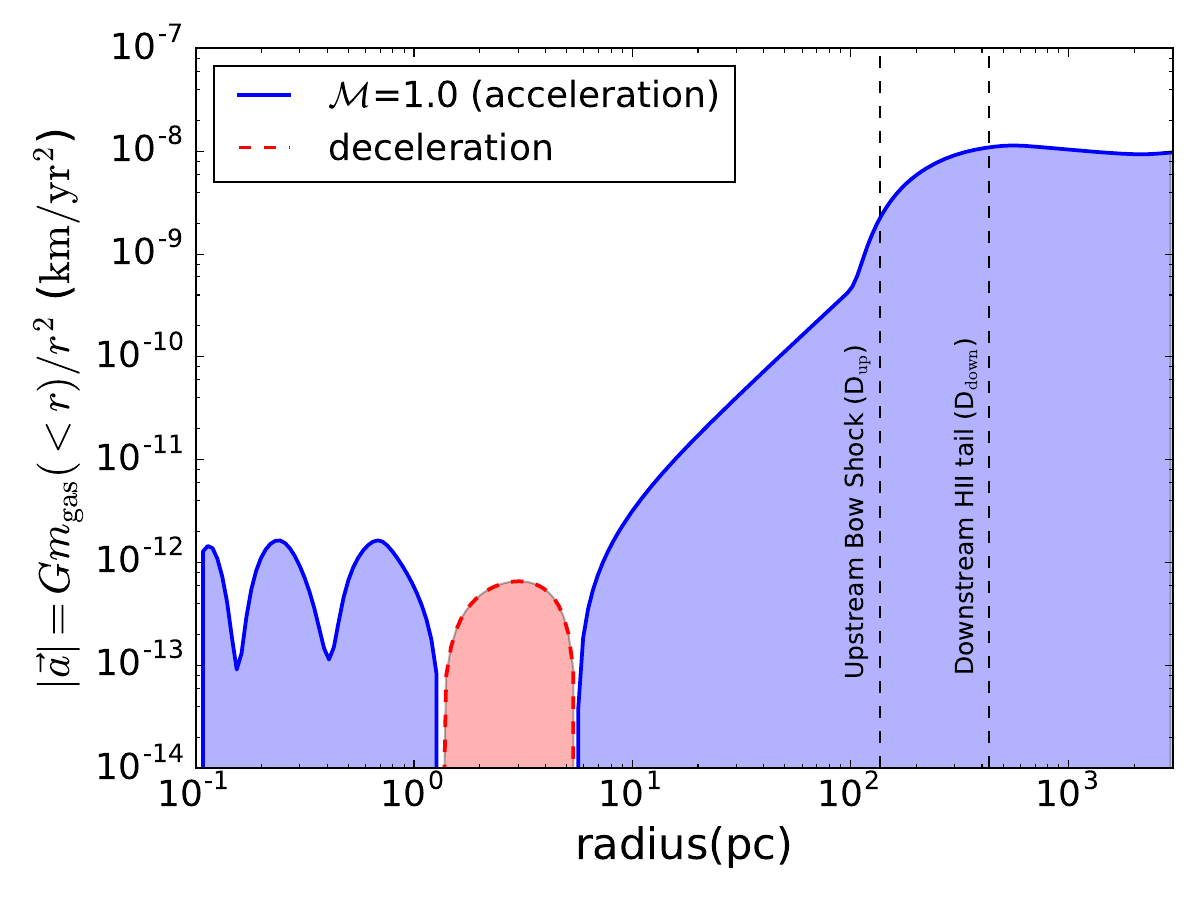}{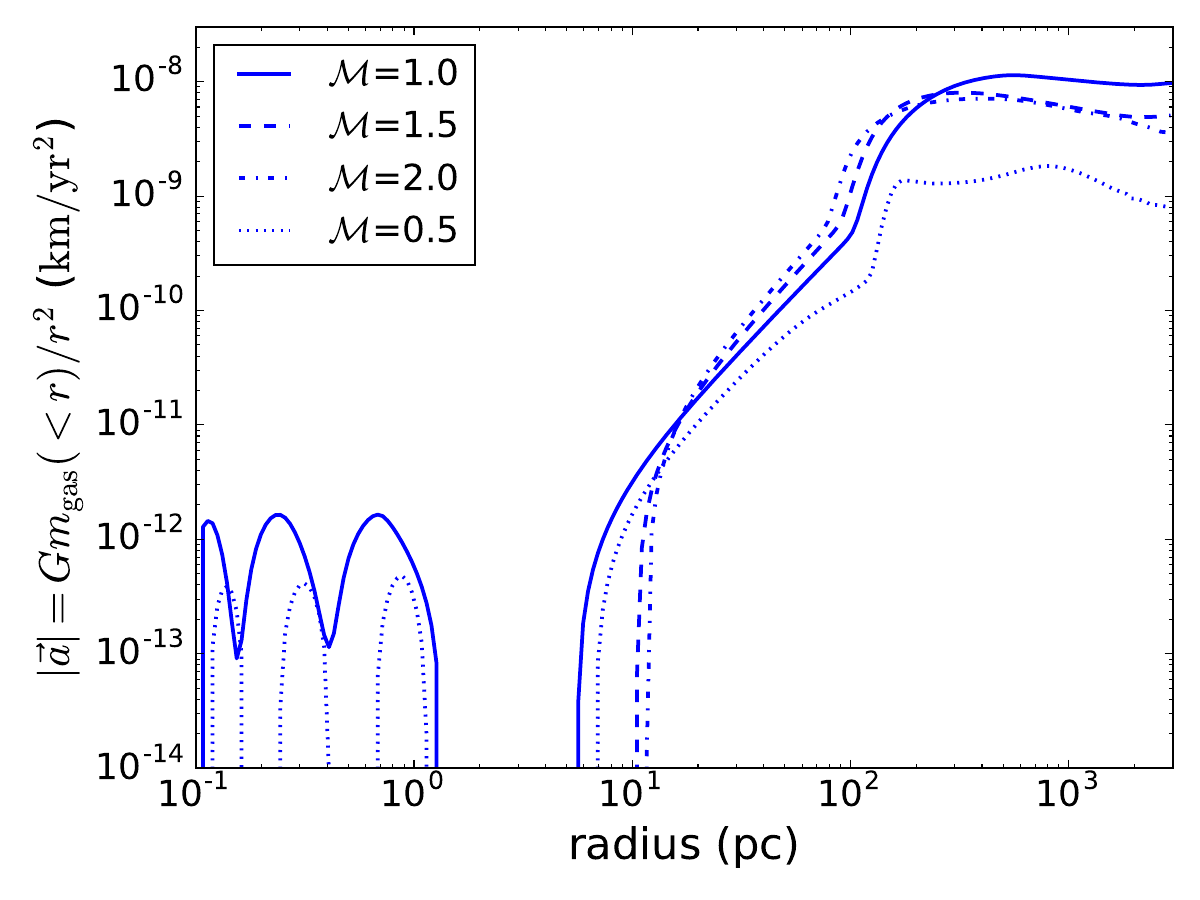}
\caption{{Left}: magnitude of the MBH acceleration along the direction of motion (either
positive or negative) contributed by the gas within polar radius
$r$ for the LR $\mach=1.0$ run. The MBH is located at the origin $r=0$.
Dashed lines mark the locations of the upstream bow shock and the
tail of the downstream \hii~region, respectively. {Right}:
magnitude of the positive component of acceleration (resulting in
MBH speed-up) contributed by the gas within $r$.
Different lines mark LR runs with $\mach$ = 0.5, 1.0, 1.5, and 2.0.}
\label{fig:delta_acc} \end{figure*}

\section{Radiation-hydrodynamic simulations}
\label{sec:method}

\subsection{Numerical Setup}
\label{sec:setup}

We run a set of 2D radiation hydrodynamic simulations using a
parallel version of the non-relativistic code ZEUS-MP
\citep{StoneN:92,Hayes:06}. Simulations are carried out in a polar
coordinate system defined by coordinates ($r$, $\theta$) and assuming
axisymmetry with respect to the MBH's direction of motion. The
extent of the computational domain is given by $r \in (0.1\,{\rm pc},
3.0\,{\rm kpc})$ and $\theta \in (0,\pi)$ and other relevant quantities are shown
in Table~\ref{table:para}. In the calculation of gas dynamics we consider
only the gravitational potential of the MBH and neglect the
self-gravity of the gas.

An MBH with mass $\mbh=10^6\,\msun$, located at the origin of the
coordinate system ($r=0$), is placed in a ``wind tunnel." In this
setup, the gas of uniform density $n_\infty=1.0\,{\rm cm}^{-3}$ and
temperature $T_\infty=10^4$\,K is
assumed to flow into the computational domain in direction $\theta=\pi$.
We evaluate the accretion rate onto the MBH by calculating the mass
flux through the inner boundary of the computational domain, defined
by the sphere with radius $r_{\rm min} = 0.1$\,pc, and convert it to MBH
luminosity as $L_{\rm bol}=0.1 \dot{M} c^2$. The SED of ionizing
radiation is described as $L_{\nu} \propto \nu^{-1.5}$ using 50
energy bins ranging from 13.6\,eV to $100$\,keV.

The composition of the H--He gas is evolved by following the species
of $\hi$, $\hii$, ${\rm He}\,{\textsc {i}}$, ${\rm He}\,{\textsc
{ii}}$, ${\rm He}\,{\textsc {iii}}$, and $e^-$. \khp{The radiation
transport is coupled with hydrodynamics and includes photo-heating,
photo-ionization, Compton heating by UV and X-ray photons from the
BH, and gas cooling \citep{ParkR:11,ParkR:12,Park:2014a,Park:2014b}.
We assume that BH radiation is the only heating source and adopt a simple analytic form of atomic cooling function $\Lambda(T)$
due to neutral and ionized H and He, which includes cooling by
recombination, collisional ionization/excitation, free-free
transitions, and di-electric recombination of ${\rm He}\,{\textsc
{ii}}$ \citep{ShapiroKang:87}.
Molecular cooling is neglected and the cooling rate is set to zero
below $T=10^4$\,K.}
The calculation of radiation transport accounts for the radiation
pressure and thus allows us to accurately capture the effects of
both energy and momentum feedback \citep[see][for implementation
details]{ParkR:12}.  Hydrodynamic and radiative transfer equations
are solved at every
time step defined as $\Delta t=$min$(\Delta_{\rm hydro}, \Delta_{\rm
chem})$, where $\Delta_{\rm hydro}$ is the hydrodynamical time step
and $\Delta_{\rm chem}$ is the time step required to calculate the
change of chemical abundances
\citep{RicottiGS:01,WhalenN:06}.

We investigate the effect of radiative feedback on the dynamical
friction wake by varying the MBH Mach number from 0.5 (subsonic)
to 2.0 (supersonic), while keeping $\mbh$, $n_\infty$, and $T_\infty$
fixed. Each simulation is run for several sound crossing timescales,
defined as $t_{\rm cross} = R_{\rm HII}/c_{s,\infty}$, which
corresponds to 46\,Myr for the size of the \hii~region defined by
equation~(\ref{eq:R_HII}). This ensures that the accretion rate
onto the MBH and density distribution of the gas in the dynamical
friction wake reach steady state. The length of each simulation is
recorded as $t_{\rm end}$ in Table~\ref{table:para}. We test numerical
convergence with the set of high (HR), medium (MR), and low resolution
(LR) runs. The HR runs are carried out with resolution of $(N_r
\times N_\theta) = (400\times128)$, MR with $(300 \times 96)$, and
LR with $(200 \times 64)$. The radial bins are logarithmically
spaced so that $\Delta r/r$ is constant everywhere on the grid while
the bins in polar angle are evenly spaced and have size $\Delta\theta
= \pi/N_{\theta}$. The outer boundary of the computational domain
is defined as inflow where $0 \le \theta \le \pi/2$ and outflow for
$\pi/2 < \theta \le \pi$.  We apply the outflow boundary conditions
at the inner domain boundary.



\subsection{Evolution of the \hii~region and overdensity wake} 
\label{sec:evolution}
 
The snapshots in Figure~\ref{fig:snapshot_m05} illustrate the
evolution of the gas density in the HR run with $\mach = 0.5$. A
low density \hii~region with $T_{\rm in} \approx 6\times10^4$\,K
forms promptly around the MBH, while the distribution of gas outside
of it reaches steady state 150\,Myr after the beginning of the
simulation. This timescale is consistent with $\sim7$ sound crossing
times for the \hii~region of the size 200\,pc. Once steady state
is achieved, the shape of the $\hii$~region shows minor deviation
from spherical symmetry.

Figure~\ref{fig:accrate} shows the accretion rate as a function of
time for the MR runs with $\mach$ = 0.5, 1.0, and 2.0. In all
simulations MBH initially exhibits a relatively high accretion rate,
which decreases as the \hii~region expands into the background
medium as a consequence of the MBH radiative feedback. The expansion
of the \hii~sphere is brought to a halt at $\sim15-30$\,Myr after
the beginning of the simulation, at which point the accretion rate
reaches a turning (minimum) point and readjusts to a steady state
value after $\sim 20-35$\,Myr. This happens first in the LR run
with $\mach$ = 2.0, followed by $\mach$ = 1.0 and 0.5. This hierarchy
of timescales is determined by the inflow rate of the gas into the
\hii~region as seen by the MBH.

Figure~\ref{fig:accrate} also illustrates the dependence $\dot{M}
\propto \mach^2$ captured by Equation~(\ref{eq:Mdot}) and applicable
in the range $1 < \mathcal{M} < 4$. This dependence of accretion
rate on the Mach number was first reported by PR13, who noted that
accretion onto the MBH is mediated by a dense shell, which forms
upstream from the MBH, at the interface of the \hii~region and
ambient gas. This dense shell and the associated bow shock are
discernible in Figure~\ref{fig:snapshot}, in the MR runs with $\mach$
= 1.0 and 2.0, and are absent for the subsonic run with $\mach$ =
0.5. The same figure shows that the \hii~region becomes more elongated
in the direction of the MBH motion with the increasing Mach number,
engulfing a larger volume in the region where the dynamical friction
wake is supposed to form. Despite this effect, a noticeable overdensity
of gas persists at the tail of the \hii~region in all simulations,
indicating that some fraction of the dynamical friction wake remains
present.

The bottom panels of Figure~\ref{fig:snapshot} illustrate the
strength of gravitational interaction between the MBH and fluid
elements in the computational domain, where we evaluate the magnitude
of the acceleration per unit volume along the direction of the BH motion
as $G\mbh\, n | \cos{\theta} | /r^2$. The net effect of a given
fluid element on the MBH depends on its location. The gas in the
upstream ($0 \leq \theta \leq \pi/2$) acts to accelerate the MBH,
while the gas in the downstream ($\pi/2 \leq \theta \leq \pi$)
decelerates it. The shade of the color indicates that the MBH most
strongly interacts with the gas in its immediate vicinity, within
a radius of a few parsecs, but this distribution appears front--back
symmetric and is thus not expected to significantly contribute to
the acceleration of the MBH. Similarly, the fluid elements perpendicular
to the MBH's line of motion at $\theta \approx \pi/2$ make a
negligible contribution. The overdensity of gas that forms at the
front and tail of the \hii~region is thus expected to contribute
the most to the net acceleration of the MBH. In the run with
$\mach$ = 0.5 this distribution is approximately front-back symmetric,
indicating a lower magnitude of acceleration. The $\mach$ = 1.0
and 2.0 runs, on the other hand, show enhanced contribution to MBH
acceleration from the front of the \hii~region, while contribution
from the tail appears less significant. In order to determine whether
the MBH accelerates or decelerates, once the steady state distribution
of gas is achieved, we integrate contributions to the dynamical
friction force from all fluid elements in the computational domain.

\subsection{Efficiency of dynamical friction} 
\label{sec:efficiency}

The left panel of Figure~\ref{fig:delta_acc} shows the steady state magnitude of the MBH acceleration (either positive or negative) due to the gas enclosed within the sphere of radius $r$ for the LR run with $\mach=1.0$. Positive acceleration is defined to be in the direction of the MBH motion, thus resulting in the speed-up of the MBH. Note that the assumption of azimuthal symmetry guarantees that the components of acceleration perpendicular to the MBH's line of motion cancel out, leaving the parallel components as the only contribution.

The magnitude of acceleration is relatively low in the immediate vicinity of the MBH ($r < 3$\,pc), consistent with the earlier observation of the front--back symmetry in this region. At $r \ga 3$\,pc the magnitude of acceleration gradually increases as the contrast between the upstream and downstream distribution of gas increases with radius. At $r \approx 100$\,pc, acceleration shows a sudden jump coinciding with the upstream overdensity and associated bow shock. The acceleration magnitude levels off beyond 400\,pc, which marks the spatial extent of the tail of the \hii~sphere. Evidently, the dynamical friction wake does not contribute to the MBH acceleration beyond this point.  

The right panel of Figure~\ref{fig:delta_acc} shows the magnitude of the positive component of MBH acceleration contributed by the gas enclosed within radius $r$. The figure illustrates that in all simulations the largest contribution to the MBH acceleration originates from the region $r \gtrsim 10$\,pc. This acceleration is net positive, thus speeding up the MBH. In all cases, the MBH deceleration due to the wake beyond the tail of the \hii~region is negligible. The acceleration originating from $r < 10$\,pc can be either positive or negative but is orders of magnitude smaller relative to the larger spatial scales and can be neglected. 

\begin{figure}[t]
\epsscale{1.15} 
\plotone{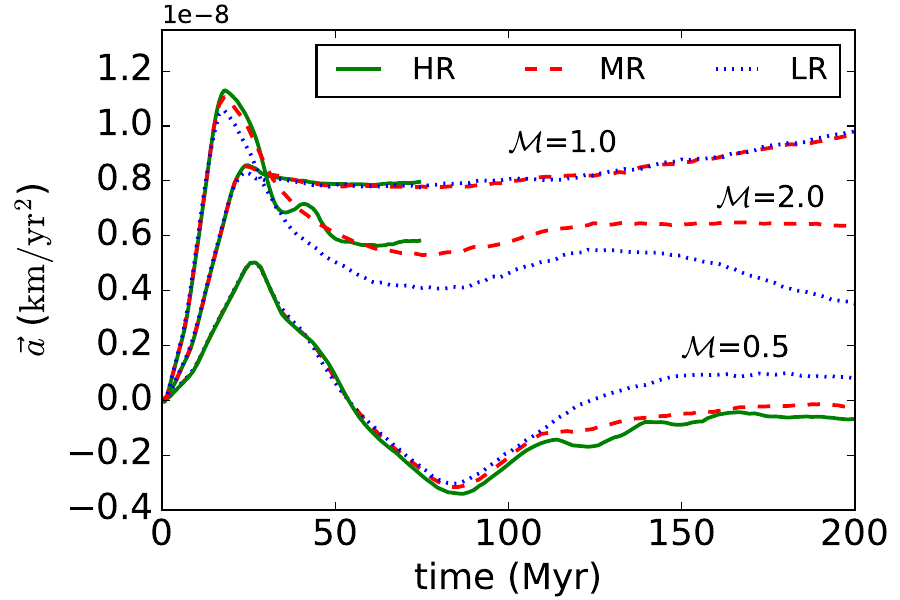} 
\caption{Evolution of the net acceleration, integrated over the
entire computational domain, for the HR (solid lines), MR (dashed),
and LR (dotted) runs with $\mach$ = 0.5, 1.0, and 2.0. Positive
acceleration implies speed-up of the MBH.}
\label{fig:cross} \end{figure}

Figure~\ref{fig:cross} shows the evolution of the net acceleration
integrated over the entire computational domain as a function of
time. In all cases acceleration initially peaks at $15-30$\,Myrs.
This is the same point in time where the accretion rate reaches
minimum (Figure~\ref{fig:accrate}), indicating a buildup of the
dense shell of gas in the upstream of the MBH, which exerts
gravitational influence and initially suppresses accretion onto the
MBH. Gravitational influence of the upstream gas shell is
counterbalanced by the gas in the downstream of the MBH, where
overdensity wake develops with a delay corresponding to a few sound
crossing times across the elongated \hii~region. Consequently, the
MBH acceleration reaches steady state with a delay of $\sim 150$\,Myr,
long after the accretion rate has settled into steady state.
An inspection of Figure~\ref{fig:cross} shows that the net acceleration
for the scenario $\mach = 1.0$ continues to gently increase even
after 200\,Myr.

For $\mach = 1.0$ and 2.0 the net end value of acceleration is
positive, indicating a speed-up of the MBH. As expected, the magnitude
of acceleration is smallest for the $\mach$ = 0.5 scenario, where
the gas flow around the MBH exhibits a large degree of front--back
symmetry. Figure~\ref{fig:cross} illustrates that the sign of net
acceleration in this case is resolution dependent, where the final
outcome of the LR run is MBH speed-up and in the MR and HR runs,
deceleration. The net accelerations in the MR and HR runs track each
other closely and differ by only $\sim 0.05\,{\rm km\,yr^{-2}}$,
indicating numerical convergence of the MR run. We therefore use
this value as a measure of error in net acceleration due to the
finite numerical resolution of our simulations.


\begin{figure}[t]
\epsscale{1.1} \plotone{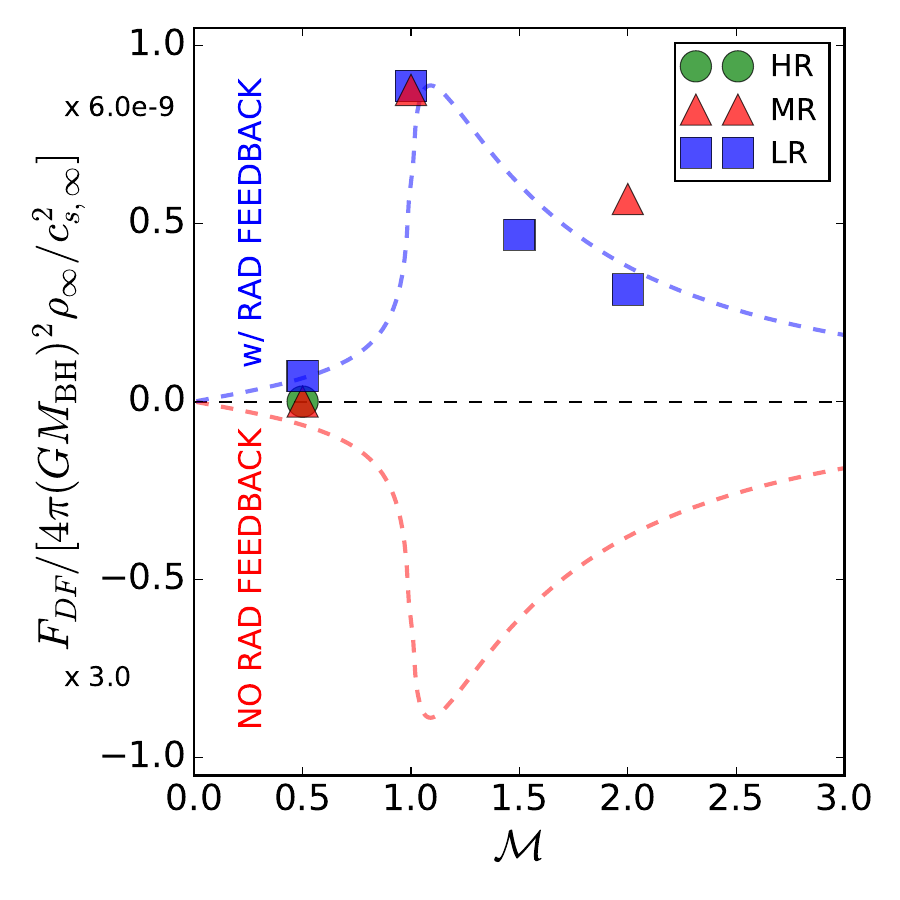}
\caption{Dynamical friction force as a function of the Mach number.
The red dashed line illustrates the magnitude of the dynamical friction
force in the absence of radiative feedback as calculated by \citet{ostriker99}.
The blue dashed line is the same as the red one but plotted with a positive
sign and arbitrarily scaled in magnitude to match data points.
Different symbols mark results of the runs with radiative feedback
at $t=201.0$\,Myr. Note the different magnitudes of the positive
and negative $y$-axis.} 
\label{fig:df_mach} \end{figure}

Figure~\ref{fig:df_mach} shows the values of the dynamical friction
force measured at the end of each simulation for different values
of $\mach$. As a comparison, the red dashed line shows the dynamical
friction force in the absence of radiative feedback, as predicted by
\citet{ostriker99} for ln\,($c_{s, \infty} t/r_{\rm min})=4$. As noted before,
the dynamical friction force calculated from simulations presented
in this study is net positive for $1.0 \le \mach < 4$ and negative for $\mach = 0.5$.
The magnitude of this force is however over $\sim 9$ orders of magnitude
lower compared to that experienced by the MBHs in absence
of radiative feedback and is therefore negligible for all values of the Mach number $\mach < 4$. 
Interestingly, the dynamical friction force in the presence of radiative
feedback appears to mirror its classical counterpart: it peaks
around $\mach \approx 1$ and decreases at larger values of the Mach
number.


\section{Discussion and conclusions} 
\label{sec:conclusions}

We investigate the efficiency of dynamical friction in presence of
radiative feedback from an MBH moving through a uniform density
gas. Ionizing radiation that emerges from the innermost parts of
the MBH's accretion flow results in the formation of the \hii~region,
which strongly affects the dynamical friction wake and renders
dynamical friction inefficient for a range of physical scenarios.
We summarize our main findings below:

\begin{itemize}
\item We identify a physical regime in which the dynamical friction
wake is likely to be fully ionized and dispersed by the MBH radiative
feedback as a regime in which the radius of the \hii~sphere is
larger than the extent of the dynamical friction wake. This condition
is fulfilled when $\mathcal{M} < 4$ and $(1+\mathcal{M}^2)\,
\mbh\,n_\infty < 10^9 \msun\,{\rm cm^{-3}}$. Outside of
this regime the formation of the \hii~region is suppressed and the
properties of the gas flow around the MBH are restored to the
classical Bondi--Hoyle--Lyttleton solution, thus restoring the effect
of dynamical friction. These criteria can be utilized as a
sub-resolution model for dynamical friction in large scale simulations
that do not resolve the scale of the \hii~region or MBH density
wake.  

\item Based on radiation hydrodynamic simulations we find that the 
net acceleration experienced by the MBHs in this regime tends to 
be positive, meaning that they speed up, contrary to the expectations
for gaseous dynamical friction in absence of radiative feedback.
This reversal happens because the dominant contribution to the MBH
acceleration comes from the dense shell of gas that forms in front
of the MBH as a consequence of the snowplow effect caused by radiation
pressure.

\item The magnitude of MBH acceleration peaks at $\mach \approx 1$
and decreases for larger values of the Mach number, similar to the
dynamical friction force in absence of radiative feedback. The
magnitude of acceleration is however negligibly small, implying
that MBHs in this regime will not significantly change their velocity
over timescales that determine their motion and properties of their
environment.  

\item Our results suggest that suppression of dynamical friction
by radiative feedback should be more severe at the lower mass end
of the MBH spectrum, because these BHs must reside in regions of
relatively high gas density in order to experience efficient dynamical
friction in the regime when $(1+\mathcal{M}^2)\, \mbh\,n_\infty >
10^9 \msun\,{\rm cm^{-3}}$.  Compounded with inefficiency of the
gas drag for lower mass objects in general this implies that
$<10^7\,M_\odot$ MBHs have fewer means to reach the centers of
merged galaxies.  

\end{itemize}

This study makes several idealized assumptions, such as the uniform
gas density and infinite background medium. In reality however,
MBHs in the aftermath of galactic mergers are likely to find
themselves immersed in inhomogeneous and clumpy medium which can
perturb their otherwise smooth orbital decay \citep{fiacconi13}.
\khp{Furthermore, the geometry and properties of the \hii~region
will be affected in scenarios where the \hii~region is not confined
by the gaseous medium. An example of this would be an \hii~region
around the MBH orbiting within the galactic gas disk. When the
radius of the \hii~region exceeds the half thickness of the disk,
the radiation can escape outside of the disk plane, making the
problem fully three-dimensional. Even in such scenarios, the proximity
of the dynamical friction wake to the MBH is likely to result in
the wake obliteration by high energy radiation. Indeed, 3D hydrodynamic
simulations that capture the dynamics of MBHs in proper galactic
setups also find a weakening of the dynamical friction force in
the presence of MBH feedback \citep{Sijacki:11, SouzaLima:2016}. The
high resolution, local simulations presented here support these
findings and provide a set of criteria, formulated in terms of the
properties of the gas and MBH, under which the wake evacuation is
efficient.}


Along similar lines, we consider an isolated MBH on a linear
trajectory, whereas the front--back asymmetry of the density wake
for a perturber on a circular orbit has been shown to cause small
differences in the dynamical friction force in absence of radiative
feedback \citep{kim07}. A system consisting of multiple MBHs would
further add to the complexity in cases when dynamical friction wakes
can  mutually affect one another \citep[e.g.,][]{KimKimS:2008}.
This type of study may require a transition from 2D to 3D simulations
in the future in order to relax the assumption of azimuthal symmetry
and properly capture the interplay of inhomogeneities, radiative
feedback, and dynamical friction. 


\acknowledgments
This work is supported in part by the National Science Foundation
under the Theoretical and Computational Astrophysics Network (TCAN)
grant AST-1333360. T.B. acknowledges support from the Research
Corporation for Science Advancement through a Cottrell Scholar
Award. T.B. is a member of the MAGNA project
(http://www.issibern.ch/teams/agnactivity) supported by the
International Space Science Institute (ISSI) in Bern, Switzerland.
Numerical simulations presented in this paper were performed using
the high-performance computing cluster PACE, administered by the
Office of Information and Technology at the Georgia Institute of
Technology.

\bibliographystyle{aasjournal} \bibliography{df}

\begin{thebibliography}{}
\expandafter\ifx\csname natexlab\endcsname\relax\def\natexlab#1{#1}\fi
\providecommand{\url}[1]{\href{#1}{#1}}

\bibitem[{{Begelman}(1978)}]{Begelman:78}
{Begelman}, M.~C. 1978, \mnras, 184, 53

\bibitem[{{Begelman}(1979)}]{Begelman:79}
---. 1979, \mnras, 187, 237

\bibitem[{{Bondi} \& {Hoyle}(1944)}]{BondiH:44}
{Bondi}, H., \& {Hoyle}, F. 1944, \mnras, 104, 273

\bibitem[{{Callegari} {et~al.}(2011){Callegari}, {Kazantzidis}, {Mayer},
  {Colpi}, {Bellovary}, {Quinn}, \& {Wadsley}}]{callegari11}
{Callegari}, S., {Kazantzidis}, S., {Mayer}, L., {et~al.} 2011, \apj, 729, 85

\bibitem[{{Callegari} {et~al.}(2009){Callegari}, {Mayer}, {Kazantzidis},
  {Colpi}, {Governato}, {Quinn}, \& {Wadsley}}]{callegari09}
{Callegari}, S., {Mayer}, L., {Kazantzidis}, S., {et~al.} 2009, \apjl, 696, L89

\bibitem[{{Chandrasekhar}(1943)}]{chandra43}
{Chandrasekhar}, S. 1943, \apj, 97, 255

\bibitem[{{Chapon} {et~al.}(2013){Chapon}, {Mayer}, \& {Teyssier}}]{chapon13}
{Chapon}, D., {Mayer}, L., \& {Teyssier}, R. 2013, \mnras, 429, 3114

\bibitem[{{Dotti} {et~al.}(2006){Dotti}, {Colpi}, \& {Haardt}}]{dotti06}
{Dotti}, M., {Colpi}, M., \& {Haardt}, F. 2006, \mnras, 367, 103

\bibitem[{{Escala} {et~al.}(2005){Escala}, {Larson}, {Coppi}, \&
  {Mardones}}]{escala05}
{Escala}, A., {Larson}, R.~B., {Coppi}, P.~S., \& {Mardones}, D. 2005, \apj,
  630, 152

\bibitem[{{Fiacconi} {et~al.}(2013){Fiacconi}, {Mayer}, {Ro{\v s}kar}, \&
  {Colpi}}]{fiacconi13}
{Fiacconi}, D., {Mayer}, L., {Ro{\v s}kar}, R., \& {Colpi}, M. 2013, \apjl,
  777, L14

\bibitem[{{Hayes} {et~al.}(2006){Hayes}, {Norman}, {Fiedler}, {Bordner}, {Li},
  {Clark}, {ud-Doula}, \& {Mac Low}}]{Hayes:06}
{Hayes}, J.~C., {Norman}, M.~L., {Fiedler}, R.~A., {et~al.} 2006, \apjs, 165,
  188

\bibitem[{{Hoyle} \& {Lyttleton}(1939)}]{HoyleL:39}
{Hoyle}, F., \& {Lyttleton}, R.~A. 1939, in Proceedings of the Cambridge
  Philosophical Society, Vol.~35, Proceedings of the Cambridge Philosophical
  Society, 405--+

\bibitem[{{Inayoshi} {et~al.}(2016){Inayoshi}, {Haiman}, \&
  {Ostriker}}]{Inayoshi:2016}
{Inayoshi}, K., {Haiman}, Z., \& {Ostriker}, J.~P. 2016, \mnras, 459, 3738

\bibitem[{{Kim} \& {Kim}(2007)}]{kim07}
{Kim}, H., \& {Kim}, W.-T. 2007, \apj, 665, 432

\bibitem[{{Kim} {et~al.}(2008){Kim}, {Kim}, \&
  {S{\'a}nchez-Salcedo}}]{KimKimS:2008}
{Kim}, H., {Kim}, W.-T., \& {S{\'a}nchez-Salcedo}, F.~J. 2008, \apjl, 679, L33

\bibitem[{{Li}(2011)}]{Li:11}
{Li}, Y. 2011, ArXiv e-prints, arXiv:1109.3442

\bibitem[{{Mayer}(2013)}]{mayer13}
{Mayer}, L. 2013, Classical and Quantum Gravity, 30, 244008

\bibitem[{{Mayer} {et~al.}(2007){Mayer}, {Kazantzidis}, {Madau}, {Colpi},
  {Quinn}, \& {Wadsley}}]{mayer07}
{Mayer}, L., {Kazantzidis}, S., {Madau}, P., {et~al.} 2007, Science, 316, 1874

\bibitem[{{Milosavljevi{\'c}} {et~al.}(2009){Milosavljevi{\'c}}, {Couch}, \&
  {Bromm}}]{MiloCB:09}
{Milosavljevi{\'c}}, M., {Couch}, S.~M., \& {Bromm}, V. 2009, \apjl, 696, L146

\bibitem[{{Osterbrock} \& {Ferland}(2006)}]{osterbrock06}
{Osterbrock}, D.~E., \& {Ferland}, G.~J. 2006, {Astrophysics of gaseous nebulae
  and active galactic nuclei}

\bibitem[{{Ostriker}(1999)}]{ostriker99}
{Ostriker}, E.~C. 1999, \apj, 513, 252

\bibitem[{{Pacucci} \& {Ferrara}(2015)}]{PacucciF:2015}
{Pacucci}, F., \& {Ferrara}, A. 2015, \mnras, 448, 104

\bibitem[{{Park} \& {Ricotti}(2011)}]{ParkR:11}
{Park}, K., \& {Ricotti}, M. 2011, \apj, 739, 2

\bibitem[{{Park} \& {Ricotti}(2012)}]{ParkR:12}
---. 2012, \apj, 747, 9

\bibitem[{{Park} \& {Ricotti}(2013)}]{ParkR:13}
---. 2013, \apj, 767, 163

\bibitem[{{Park} {et~al.}(2014{\natexlab{a}}){Park}, {Ricotti}, {Di Matteo}, \&
  {Reynolds}}]{Park:2014a}
{Park}, K., {Ricotti}, M., {Di Matteo}, T., \& {Reynolds}, C.~S.
  2014{\natexlab{a}}, \mnras, 437, 2856

\bibitem[{{Park} {et~al.}(2014{\natexlab{b}}){Park}, {Ricotti}, {Di Matteo}, \&
  {Reynolds}}]{Park:2014b}
---. 2014{\natexlab{b}}, \mnras, 445, 2325

\bibitem[{{Park} {et~al.}(2016){Park}, {Ricotti}, {Natarajan},
  {Bogdanovi{\'c}}, \& {Wise}}]{Park:2016}
{Park}, K., {Ricotti}, M., {Natarajan}, P., {Bogdanovi{\'c}}, T., \& {Wise},
  J.~H. 2016, \apj, 818, 184

\bibitem[{{Ricotti} {et~al.}(2001){Ricotti}, {Gnedin}, \&
  {Shull}}]{RicottiGS:01}
{Ricotti}, M., {Gnedin}, N.~Y., \& {Shull}, J.~M. 2001, \apj, 560, 580

\bibitem[{{Shakura} \& {Sunyaev}(1973)}]{ShakuraS:73}
{Shakura}, N.~I., \& {Sunyaev}, R.~A. 1973, \aap, 24, 337

\bibitem[{{Shapiro} \& {Kang}(1987)}]{ShapiroKang:87}
{Shapiro}, P.~R., \& {Kang}, H. 1987, \apj, 318, 32

\bibitem[{{Sijacki} {et~al.}(2011){Sijacki}, {Springel}, \&
  {Haehnelt}}]{Sijacki:11}
{Sijacki}, D., {Springel}, V., \& {Haehnelt}, M.~G. 2011, \mnras, 414, 3656

\bibitem[{{Souza Lima} {et~al.}(2016){Souza Lima}, {Mayer}, {Capelo}, \&
  {Bellovary}}]{SouzaLima:2016}
{Souza Lima}, R., {Mayer}, L., {Capelo}, P.~R., \& {Bellovary}, J.~M. 2016,
  ArXiv e-prints, arXiv:1610.01600

\bibitem[{{Stone} \& {Norman}(1992)}]{StoneN:92}
{Stone}, J.~M., \& {Norman}, M.~L. 1992, \apjs, 80, 753

\bibitem[{{Whalen} \& {Norman}(2006)}]{WhalenN:06}
{Whalen}, D., \& {Norman}, M.~L. 2006, \apjs, 162, 281

\end{thebibliography}


\begin{thebibliography}{}
\expandafter\ifx\csname natexlab\endcsname\relax\def\natexlab#1{#1}\fi
\providecommand{\url}[1]{\href{#1}{#1}}
\providecommand{\dodoi}[1]{doi:~\href{http://doi.org/#1}{\nolinkurl{#1}}}
\providecommand{\doeprint}[1]{\href{http://ascl.net/#1}{\nolinkurl{http://ascl.net/#1}}}
\providecommand{\doarXiv}[1]{\href{https://arxiv.org/abs/#1}{\nolinkurl{https://arxiv.org/abs/#1}}}

\bibitem[{{Gruzinov} {et~al.}(2020){Gruzinov}, {Levin}, \&
  {Matzner}}]{gruzinov19}
{Gruzinov}, A., {Levin}, Y., \& {Matzner}, C.~D. 2020, \mnras, 492, 2755,
  \dodoi{10.1093/mnras/staa013}

\bibitem[{{Ostriker}(1999)}]{ostriker99}
{Ostriker}, E.~C. 1999, \apj, 513, 252, \dodoi{10.1086/306858}

\bibitem[{{Park} \& {Ricotti}(2013)}]{park13}
{Park}, K., \& {Ricotti}, M. 2013, \apj, 767, 163,
  \dodoi{10.1088/0004-637X/767/2/163}

\end{thebibliography}

\label{lastpage} 
\end{document}